\documentstyle[12pt]{article}  
\def\beq{\begin{equation}}  
\def\eeq{\end{equation}}  
\def\bea{\begin{eqnarray}}  
\def\eea{\end{eqnarray}}  
\def\bq{\begin{quote}}  
\def\eq{\end{quote}}  
  
\def\beqa{\begin{eqnarray}}  
\def\eeqa{\end{eqnarray}}  
\def\be{\begin{equation}}  
\def\ee{\end{equation}}  
\def\beq{\begin{equation}}  
\def\eeq{\end{equation}}

\def\bi{\begin{itemize}}  
\def\ei{\end{itemize}}

\newcommand{\barre}[1]{%
        \setbox1=\hbox{$#1$} \dimen2=\ht1 \dimen3=\dp1 \dimen4=\wd1  
        \setbox2=\hbox{\sl /}  
        \dimen1=\wd1 \advance\dimen1 by -\wd2 \divide\dimen1 by 2  
        \advance\dimen1 by \wd2 \advance\dimen1 by 0.4pt  
        \setbox3=\hbox to \wd1{\hss \box1 \kern -\dimen1 \box2\hss}  
        \ht3=\dimen2 \dp3=\dimen3 \wd3=\dimen4  
        \box3  
        }  
  
\begin{document}  
\pagestyle{empty}  
\begin{flushright}  
                     hep-th/0002164  
\end{flushright}  
\vskip 2cm    
  
\begin{center}  
{\huge A Comment on Self-Tuning and Vanishing Cosmological Constant in   
the Brane World\\}  
\vspace*{5mm} \vspace*{1cm}   
\end{center}  
\vspace*{5mm} \noindent  
\vskip 0.5cm  
\centerline{\bf Stefan F\"orste${}^{1}$, Zygmunt Lalak${}^{1,2}$,
St\'ephane Lavignac${}^{1,3}$ and Hans Peter Nilles${}^{1}$
}
\vskip 1cm
\centerline{\em ${}^{1}$Physikalisches Institut, Universit\"at Bonn}
\centerline{\em Nussallee 12, D-53115 Bonn, Germany}
\vskip 0.3cm
\centerline{\em ${}^{2}$Institute of Theoretical Physics}
\centerline{\em Warsaw University, Poland}
\vskip 0.3cm
\centerline{\em ${}^{3}$Service de Physique Th\'eorique, CEA-Saclay}
\centerline{\em F-91191 Gif-sur-Yvette C\'edex, France}
\vskip2cm 
  
\centerline{\bf Abstract}  
\vskip .3cm  
In this note we elaborate on various five dimensional contributions to the
effective 4D cosmological constant in brane systems. In
solutions with vanishing   
5D cosmological constant we describe a non-local mechanism of  
cancellation of vacuum energy between the brane and the  
singularities. We comment on a hidden fine tuning which is implied by  
this observation.

  
\newpage  
  
\setcounter{page}{1} \pagestyle{plain}

One of the main issues in the `Brane World' scheme of the superunification   
of gauge forces with gravity is the question of    
having the observable sector (containing the Standard Model degrees of
freedom)   
confined to a lower dimensional brane embedded into higher-dimensional bulk   
without violating experimental bounds on corrections to the precisely   
measured parameters of the Standard Model and Newton's law of
gravity.    
In this context the prominent observable which needs to be taken care of   
is the four dimensional cosmological constant, known to be   
close to zero at present. In four dimensions the vanishing cosmological   
constant, or in fact vanishing vacuum expectation value of the   
energy-momentum tensor, is necessary to obtain Minkowski space as the
solution    
of the 4d Einstein equations. This leads one to consider vacuum   
configurations in higher dimensions which have four dimensional part of the   
metric in the form of the Minkowski metric multiplied by the warp factor   
which depends on transverse coordinates only. The simplest setup   
corresponds to just a single transverse coordinate, with the line element   
\beq \label{ansatz} 
ds^2 = e^{2 A(x^5)} \eta_{\mu \nu} dx^\mu dx^\nu +
\left(dx^5\right)^2 .\eeq   
  
The sources for such configurations are assumed to be located on    
a number of four dimensional branes, one of which represents the observable   
gauge sector. It is important to note that although the observable gauge   
interactions are strictly confined to the 3-brane, the gravity and moduli   
fields permeate the whole space, effectively connecting the walls in a   
nontrivial way. This implies that every particle localized  on the wall   
feels sources of gravitational forces which are located all over the bulk.   
In some cases the influence of the remote sources will be suppressed, like   
in the case of the exponentially falling off graviton wave function in ref.   
\cite{RS2}, which would effectively restrict the relevant   
gravitational sources to the thin layer around the brane, but sometimes   
the suppression would be so mild, that the influence of the whole bulk  
contribution will be highly relevant. Thus in any case, even though the   
gauge forces are restricted to the branes, the gravitational sector has to   
be completely integrated out when going to the effective four dimensional   
theory. In particular, this implies that when one computes the effective   
four dimensional energy density, or four dimensional vacuum pressure, one   
has to integrate over the whole causally accessible portion of the   
transverse space. The issue of exact solutions to Einstein equations   
with Minkowski-space 4d foliations and vanishing 4d vacuum energy and   
pressure can be studied in detail in the quasi-stringy setup of five   
dimensional dilaton gravity. Models of this type, with and without   
singularities located in transverse space, have been recently studied in a  
series of papers \cite{RS2,dim,kach1,kach2,wolf} (and references
therein). However, the 
problem of precisely how the cancellation of various contributions to
the vacuum energy and pressure occurs in these models has not been
explicitly addressed. The details of this    
mechanism are also important in view of the question about the amount of   
fine-tuning between contributions coming from spatially disconnected branes   
required to achieve vanishing 4d parameters. We believe that this
sheds more light on the recently advertised miraculous self-tuning
mechanism of \cite{dim},\cite{kach1}. In fact, the way we see it the
self-tuning is a fine-tuning in disguise.         
  
In this  note we shall discuss the details of the nonlocal
cancellations which occur in the five dimensional dilaton gravity
models with various number of branes, with and without bulk
cosmological constant. The 5d action  we consider is   
\beq \label{genericaction}  
S_5 = \int d^5x \sqrt{-g} ( R - \frac{4}{3} (\partial \phi )^2 - \Lambda  
e^{a \phi} ) + \int d^5x \sqrt{-g_4} ( - f_i (\phi) \delta \left(x^5 -
  x^5 _i\right))    
\eeq  
where $f_i = V_i e^{b_i \phi } $ and the index $i$ counts the branes.   
The corresponding Einstein equations are   
\beq  
\sqrt{-g} ( G_{MN} - \frac{4}{3} \partial_M \phi \partial_N \phi +  
\frac{2}{3} ( \partial \phi )^2  g_{MN} + \frac{1}{2} \Lambda e^{a
  \phi}  
g_{MN}) + \frac{1}{2} \sqrt{-g_4} f_i \delta \left(x^5-x^5 _i\right)
g_{\mu \nu}  
\delta^{\mu}_{M}  
\delta^{\nu}_{N} = 0  
\eeq  
and the dilaton equation of motion is   
\beq  
-\Lambda e^{a \phi} a \sqrt{-g} - \sqrt{-g_4}  
\delta\left(x^5 - x_i ^5\right) \partial_{x^5} f_i  
+ \frac{8}{3} \partial_M ( \sqrt{-g} g^{MN} \partial_N \phi ) = 0,   
\eeq   
where $M,N = 1, \ldots , 5$ and $\mu , \nu = 1, \dots , 4$.  
The ansatz (\ref{ansatz}) implies that the 4d cosmological constant
should vanish. In the known solutions with non-vanishing 5d constant
$\Lambda $ the vacuum energy has contributions from the bulk
integration and at the sources. Performing the bulk integral one obtains
boundary terms which locally cancel the vacuum energy at the branes.  
However, for vanishing $\Lambda $ the situation is different as we
will explain in what follows.

\vskip .3cm

Our claim ist that the effective four dimensional cosmological  
constant vanishes only when one conjectures
contributions to the   
vacuum energy being located at the singularities. 
To show in detail  how the vacuum energy located at the singularities
cancels    
the contribution from the brane\footnote{It was independently observed
  in \cite{youm}  
that taking only the contribution from the brane at the origin gives a
nonvanishing result.} 
we need to resolve the singularity. The easiest way to do this is to put  
additional sources there such that Einstein's equation is satisfied  
everywhere. Therefore we supplement the action of \cite{kach1} by  
additional source terms, and (\ref{genericaction}) takes the form  
\begin{eqnarray}  
S & = & \int d^5 x \sqrt{-G}\left[ R-\frac{4}{3}\left(\nabla  
    \phi\right)^2\right] \nonumber \\   
  & &  - \int d^4x\sqrt{-g}V{e^{b\phi}}_{|x^5 =0} \nonumber \\  
  & &  -  \int d^4x\sqrt{-g}V_+{e^{b_+\phi}}_{|x^5 =x_+}\nonumber\\  
  & &  - \int d^4x\sqrt{-g}V_-{e^{b_-\phi}}_{|x^5 =x_-}      
\end{eqnarray}  
where the indices $\pm$ refer to the singularities at $x^5 \, ^> _<\, 0$.   
This modifies equations (2.7) and (2.9) of \cite{kach1}  
correspondingly. Now let us consider in more detail how this  
influences solution {\it (I)} of \cite{kach1}. Before discussing the
effect of   
the singularities let us recall formulae which we  
are going to use in the subsequent considerations.   
First of all solution {\it (I)} is obtained with the ansatz   
$A^\prime = \mp \frac{1}{3}\phi^\prime$ for $x^5 >0$ ($x^5 < 0$), and  
the prime denotes derivative with respect to $x^5$.  
The solution for $\phi $ reads  
\begin{equation}  
\phi \left( x^5\right) =\left\{  
\begin{array}{l l}  
\frac{3}{4}\log\left|\frac{4}{3}x^5 + c_1\right| + d_1, & x^5 < 0\\  
-\frac{3}{4}\log\left|\frac{4}{3}x^5 + c_2\right| + d_2, & x^5 >0 ,  
\end{array} \right.  
\end{equation}  
where the $c_i , d_i $ are integration constants. In the following  
$c_1$ is taken to be positive and $c_2$ negative.  
Continuity at $x^5 =0$ is used to eliminate $d_2$. Boundary conditions  
on the first derivatives  
at $x^5 =0 $  give equations which can be solved for the $c_i$,  
\begin{eqnarray}  
\frac{2}{c_2} & = &\left[ -\frac{3b}{8} -\frac{1}{2}\right] V e^{bd_1}  
\left| c_1\right|^{\frac{3}{4}b} \label{tse2} \\  
 \frac{2}{c_1} & = &\left[ -\frac{3b}{8} +\frac{1}{2}\right] V e^{bd_1}  
\left| c_1\right|^{\frac{3}{4}b} . \label{tse1}   
\end{eqnarray}  
Hence, the solution does exist for any value of $V$ and $b$ and has one  
undetermined integration constant.  
  
Now, we want to investigate how putting additional sources at the  
singularities will modify this observation of self-tuning.   
By resolving the singularities with additional source terms we get two  
more boundary conditions at those points ($x_+ = -\frac{3}{4}c_2$,
$x_-=-\frac{3}{4}c_1$),  
\begin{equation}\label{fine1}  
\frac{8}{3}\left(\phi^\prime\left(x_\pm + 0\right) -\phi^\prime\left(x_\pm  
    -0\right)\right) = b_\pm V_\pm e^{b_\pm \phi\left( x_\pm\right)}  
\end{equation}  
and  
\begin{equation}\label{fine2}  
3\alpha_\mp\left(\phi^\prime\left(x_\pm + 0\right) -\phi^\prime\left(x_\pm  
    -0\right)\right) =  -\frac{1}{2}V_\pm e^{b_\pm \phi\left( x_\pm\right)}  
\end{equation}  
with $\alpha_\mp = \mp \frac{1}{3}$.  
In order to make sense out of (\ref{fine1}),(\ref{fine2}) $\phi $  
needs to be continued beyond the singularities. There are two (perhaps  
equivalent) ways of doing that: (a) periodic continuation of the  
solution or (b) cutting off the fifth direction by defining $|x-x_\pm| =  
0 $ for $x\, ^> _< \, x_\pm$ .  
We will follow the second option (b). Technically this means that for  
$x^5 > 0$ ($x^5 < 0$) we  
drop the first (second) terms on the left hand sides   
of (\ref{fine1}), (\ref{fine2}).   
We obtain the following conditions  
\begin{equation} \label{bc1}  
b_\pm = \pm \frac{4}{3}  
\end{equation}  
and  
\begin{equation} \label{bc2}  
V_- e^{-\frac{4}{3} d_1} =  
V_+ e^{\frac{4}{3} d_2} = -2 .  
\end{equation}  
Before  
plugging in the explicit value of $d_2$ we write down the contribution  
of the singularities to the vacuum energy. There are delta-peaked  
terms in the bulk Lagrangian and the additional source terms at the  
singularities. Adding this up the singularities give the following  
contribution to the four dimensional energy density  
\begin{equation}  
E_+ + E_- =-\frac{1}{3} \left({V_+ e^{4A + b_+\phi}}_{|x^5 = x_+} +  
{V_-e^{4A+b_-\phi}}_{|x^5=x_-}\right) .  
\end{equation}  
To be specific we choose $A = \frac{1}{3}\phi$ for $x^5 <  
0$. Requiring continuity at zero gives  
$A= -\frac{1}{3}\phi +\frac{1}{4}\log \left|\frac{c_1}{c_2}\right| +  
\frac{1}{3}\left( d_1 + d_2\right)$ for $x_5 > 0$.  
With (\ref{bc1}) and (\ref{bc2}) and formulae (\ref{tse2}) and  
(\ref{tse1}) we get  
\begin{equation} \label{singvev}  
E_+ + E_- = \frac{2}{3}e^{\frac{4}{3}d_1}  
  \left(\left|\frac{c_1}{c_2}\right|+1\right) = \frac{2}{3}e^{\frac{4}{3}d_1}  
  \frac{8}{4 - 3b} .  
\end{equation}  
The contribution at zero is found to be  
\begin{equation}  
E_0 = -\frac{1}{3}V{e^{4A +b\phi}}_{|x^5 =0}\,  .
\end{equation}  
Using (\ref{tse1}) one finds  
\begin{equation} \label{zerovev}  
E_0 = -\frac{2}{3} \frac{8}{4-3b}e^{\frac{4}{3}d_1} .  
\end{equation}  
Adding up (\ref{singvev}) and (\ref{zerovev}) we obtain for the 4d  
effective cosmological constant  
\begin{equation} \label{hurra}  
\Lambda_{4d} = E_+ + E_- + E_0 = 0.  
\end{equation}  
This shows that it is important to take into account contributions  
from singularities to get vanishing effective cosmological constant.  
  
A remark on self- versus fine-tuning is in order. Our conditions  
(\ref{bc1}), (\ref{bc2}) clearly impose fine-tuning on the parameters  
of the sources at the singularities. 
The hope of the authors of  
\cite{kach1}, \cite{kach2} is that a different way of resolving the  
singularities will   
automatically give the desired contributions to the cosmological  
constant. One could imagine that at the singularities new light  
degrees of freedom appear which adjust their vev such that  
(\ref{hurra}) holds. To find a specific example where one can see the  
detailed dynamics underlying the self tuning would certainly be very  
interesting.

In what follows we compute the vacuum energy for the solution {\it  
  (II)} of ref.\ \cite{kach1}  
\begin{equation}  
\phi \left( x^5\right) =\left\{  
\begin{array}{l l}  
 \pm \frac{3}{4}\log\left|\frac{4}{3}x^5 + c \right| + d, & x^5 < 0\\  
\pm \frac{3}{4}\log\left|\frac{4}{3}x^5 - c \right| + d, & x^5 >0 .  
\end{array} \right .  
\end{equation}  
 In the following  
$c$ is taken to be a positive constant. The remaining parameters 
are $b = \mp \frac{4}{3}$ and $V_0 = 4 e^{\pm \frac{4}{3} d} $. With
this choice the solution has two singularities, one on each side of
the brane.  
If one neglects possible contributions from singularities then one obtains 
after the calculation analogous to the one in the previous example the
vacuum energy  
\beq
E=E_0 = - \frac{1}{3} V_0 \neq 0.
\eeq
This is obviously nonzero for a nonzero brane tension $V_0$. The trouble is 
that, again, the solution does not fulfill the Einstein equations at the 
positions of singularities, so is not a global solution. The point is that 
the sources supporting the singularities of the Einstein tensor, and that of 
the gradient energy of the dilaton, are missing. The simplest way to repair 
the solution is to paste in suitable sources at singularities. Even without worrying about the dynamical origin of these sources, one can easily find out 
the total contribution from these sources which is needed to repair the solution and to make the vacuum energy vanish. Straightforward calculation shows that 
the total vacuum energy including contributions from singularities (and
corresponding sources) is  
\beq
E=E_0 + E_+ + E_- = E_0 - \frac{1}{3}  e^{4 A(\frac{3}{4} c )} V_+ e^{b
\phi (\frac{3}{4} c )}  - \frac{1}{3}  e^{4 A(-\frac{3}{4} c )} V_- e^{b
\phi (-\frac{3}{4} c )}
\eeq
where one finds directly from the solution that 
$V_+ = V_- = - \frac{1}{2} V_0$. Again, the total vacuum energy vanishes
when one includes the total (sources + singularities in the fields)
contribution at each singularity. This means that one needs to cut-off
the space at singularities by putting a stiff (infinite) potential
wall at each singularity, or alternatively one might imagine repeating
the whole module consisting of the original brane and two
singularities with sources along the fifth dimension. 
  
Next we would like to comment on orbifold examples. First, let us 
divide a line by $Z_2$. Then the acceptable solution must be symmetric around 
the origin. Since we do not want singularities to appear we take
$\phi=\frac{3}{4} \log (\frac{4}{3}|x^5| + c) + d$ with positive $c$. 
When one computes vacuum 
energy, then one obtains a nonzero, but finite, result. This solution 
is a valid solution of equations of motion everywhere, hence one would 
think that one obtains a consistent example of the metric which admits 
Minkowski-type foliation and gives a nonzero vacuum energy. 
However, the resolution of the puzzle comes from the observation that 
the effective 
four dimensional Planck scale which is proportional to the integral of 
the warp factor  $\sim \sqrt{ \frac{4}{3} |x^5| + c}$ diverges\footnote{
The graviton zero mode, which is actually proportional to the warp factor,
 $\sqrt{ \frac{4}{3} |x^5| + c}$ is not normalisable.}. Thus gravitational 
degrees of freedom become frozen and effectively 
gravity decouples from the physics on the brane.

The situation changes when one considers dilaton gravity on the orbifold 
$S^1 /Z_2$ extending between $-\pi \rho$ and $\pi \rho$. There the
above nonsingular solution can be extended to the solution on this
orbifold if one puts on the orbifold plane at $\pi \rho$ a system
which conspires in such a way as  
to produce there the brane tension $V_{\pi \rho} = - V_0$. The vacuum 
energy of such an orbifold vanishes due to cancellation between
contributions from the two branes. However, the correlation between
brane tensions on spatially separated branes must be considered to be
a fine-tuning, similar to that in the Randall-Sundrum model \cite{RS2}. 
We want to point out that the presence of the dynamical dilaton does not 
decrease the degree of fine-tuning with respect to the Randall-Sundrum model. 
The important result due to the dilaton is the softening of the dependence 
of the warp factor, and consequently of the graviton zero mode, on the fifth 
coordinate. This dependence changes from the exponential fall-off to the 
mild fractional power-law dependence. The result is that one cannot naturally 
produce the large hierarchy of mass scales in these quasi-stringy models.

\vskip .5 cm 
  
We have shown that in examples where the 5d cosmological constant  
vanishes there is a non local mechanism of cancellation between the  
vacuum energy at the brane and at the singularities. (For the low  
energy four dimensional observer this looks like a vanishing vacuum  
energy at the brane.) We have argued that this mechanism leads to a  
hidden fine-tuning even for the self-tuning brane solution. The  
self-tuning feature can survive only when one finds some dynamical  
mechanism by which the vacuum energy at the singularity adjusts its  
value in such a way that it cancels the contribution from the Standard Model  
brane. Without the knowledge of such a mechanism the self-tuning brane  
solution seems qualitatively quite similar to the fine tuned orbifold  
model we discussed above. For this solution it  
is much simpler to calculate corrections to Newton's law as  
there are no subtleties due to a degenerate metric. Work in that  
direction is in progress.  
  
\vskip 1cm  
  
\noindent {\bf Acknowledgments}

\noindent We thank Rados\l aw Matyszkiewicz for useful discussions
and Shamit Kachru for correspondence on  
\cite{kach1} and \cite{kach2}.   

\noindent This work has been supported by TMR programs
ERBFMRX--CT96--0045 and CT96--0090.
Z.L. is additionaly supported
by the Polish Committee for Scientific Research grant 2 P03B 05216(99-2000).

\end{document}